  \documentclass[aps,prl,twocolumn,reprint,preprintnumbers,floatfix,nofootinbib]{revtex4}

\usepackage{amsmath}
\usepackage{amssymb}
\usepackage{tikz}
\usepackage{graphicx}
\usepackage{dcolumn}
\usepackage{bm}
\usepackage{marvosym}
\usepackage{tabularx}
\usepackage{epstopdf}
\usepackage[normalem]{ulem}

\usepackage{color}

\newcommand{\beq}{\begin{equation}}
\newcommand{\eeq}{\end{equation}}

\newcommand{\rmv}[1]{\textcolor{red}{ }}

\begin{document}
\title{Discontinuous transition from direct to inverse cascade in three-dimensional turbulence\footnote{Postprint version of the manuscript published in Phys. Rev. Lett. {\bf 118}, 164501 (2017)}}
\author{Ganapati Sahoo$^1$, Alexandros Alexakis$^2$ and Luca Biferale$^1$  }
\affiliation{$^1$ Department of Physics and INFN, University of Rome 'Tor Vergata', Via della Ricerca Scientifica 1, 00133 Rome, Italy.}
\affiliation{$^2$ Laboratoire de Physique Statistique, \'Ecole Normale Sup\'erieure, CNRS, Universit\'e Pierre et Mari\'e Curie, Universit\'e Paris Diderot, 24 rue Lhomond, 75005 Paris, France.}

\begin{abstract}
Inviscid invariants of flow equations are crucial in determining the direction
of the turbulent energy cascade.  In this work we investigate a variant of the
three-dimensional  Navier-Stokes equations that shares exactly the same ideal
invariants (energy and helicity) and the same symmetries (under rotations,
reflections and scale transforms) as the original equations. It is demonstrated
that the examined system displays a change in the direction of the energy
cascade when varying the value of a free parameter which controls the relative
weights of the triadic interactions between different helical Fourier
modes.  The transition from a forward to inverse cascade is shown to occur at a
critical point in a discontinuous manner with diverging fluctuations  close to
criticality.  Our work thus supports the observation that  purely isotropic and
three-dimensional flow configurations can support inverse energy transfer when
interactions are altered and that inside all turbulent flows there is a
competition among forward and backward transfer mechanisms which might lead to
multiple energy-containing turbulent states.
\end{abstract}

\maketitle

 In turbulence the energy cascade direction determines the macroscopic
 properties of the flow, leading to a finite energy dissipation rate in the
 case of a forward cascade (from large to small scales) or to the formation of
 a condensate in the case of an inverse cascade (from small to large scales)
 \cite{Frisch}.  It has been long thought that the cascade direction is
 determined by  the dimensionality and by the ideal invariants of the flow.
 Two-dimensional turbulence possesses two sign definite invariants, the energy
 and the enstrophy.  Energy is transferred backward to larger scales while
 enstrophy is transferred forward to the small scales.  In 3D turbulence,
 energy is sign definite, while the second invariant, the helicity, is sign
 indefinite.  As a result,  helicity does not impose any local or global
 constraints and it is an empirical fact that in 3D turbulent flows both energy
 and helicity are transferred to small scales
 \cite{brissaud1973helicity,chen2003joint}. 

Other systems develop a more complex phenomenology; e.g.,  flows in thin
layers, in a stratified medium,  in the presence of rotation or of magnetic
field show a quasi-2D behavior \cite{Smith1999transfer,celani2010turbulence,
alexakis2011two, pouquet2013geophysical,marino2013inverse,
deusebio2014dimensional,seshasayanan2014on,
seshasayanan2016critical,Benavides2017,Biferale2016} and display a
bidirectional split energy cascade: part of the energy goes towards small
scales (as in 3D) and part  to the large scales (as in pure 2D flows).  This
phenomenon has also been observed in recent experiments
\cite{xia2011upscaling,Yarom2013,campagne2014direct} and in atmospheric
measurements \cite{huricanes}.  The reason for the appearance of an inverse
energy flux is ascribed to the presence of (resonant) waves or of geometric
confinement that favor the enhancement of quasi-2D Fourier interactions over
the 3D background. 

\begin{figure}[!ht]                                                             
  \includegraphics[width=0.49\textwidth]{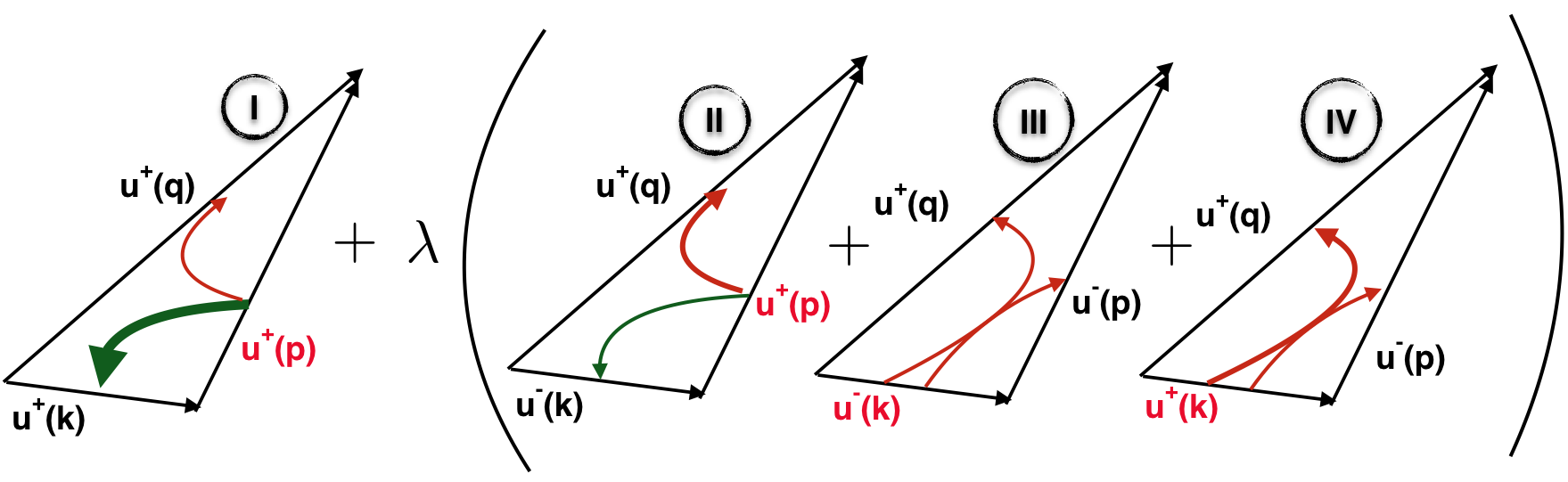}               
   \caption{Sketch of the                                        
four classes of the helical-Fourier       
decomposition of NSE. Green (red) lines describe the backward (forward) energy  
transfer from the most unstable mode \cite{waleffe1992nature}. The thicker line 
corresponds to the dominant term.}                                              
   \label{fig:triad}                                                            
 \end{figure}

In this work we study a model system for which the interactions in the
Navier-Stokes equations (NSE) are enhanced or suppressed  in a controlled way
without reducing the number of degrees of freedom, altering the inviscid
invariants or breaking any of the symmetries of the NSE.  Our study is based on
the  helical decomposition
\cite{craya1958contributiona,herring1982comparative,Lesieur72,waleffe1992nature}
of the velocity field ${\bf u }$, that in terms of its Fourier modes
$\tilde{\bf u}_{\bf k}$ is written as $\tilde{\bf u}_{\bf k}=
\tilde{u}_{\bf k}^+ {\bf h}_{\bf k}^+ + \tilde{u}_{\bf k}^- {\bf h}_{\bf k}^- $,
where ${\bf h}_{\bf k}^\pm$ are the eigenvectors of the curl operator $i {\bf
k} \times {\bf h}_{\bf k}^\pm  = \pm k {\bf h}_{\bf k}^\pm  $.  In real space the
velocity field is written as ${\bf u}={\bf u}^+ + {\bf u}^-$ where ${\bf
u}^\pm$ is the velocity field whose Fourier transform  is projected to the
${\bf h}^\pm$ base.  It is easy to realize that, in terms of the helical
decomposition, the nonlinear term of the 3D NSE  is split in 4 (8 by considering
the obvious symmetry that changes the sign of all helical modes) possible
classes of helical interactions, corresponding to triads of helical Fourier
modes, $( \tilde{u}_{\bf k}^\pm,\tilde{u}_{\bf q}^\pm,\tilde{u}_{\bf p}^\pm)$,
as depicted {by the four triadic families} in Fig.~\ref{fig:triad}.
In our simulations we change the relative
weight among homochiral triads (Class I) and all the others by introducing a
factor $0 \le \lambda \le 1$ in the nonlinear evolution. We show that
by using this weighting protocol the turbulent evolution displays a sharp
transition, for a critical value $\lambda_c$, from forward to backward energy
transfer but still  keeping the dynamics fully three-dimensional, isotropic,
and parity invariant. It
was shown in Ref. \cite{waleffe1992nature} that a generic single homochiral triad (from Class I in
Fig. \ref{fig:triad}) always leads to an excess of energy transfer to large
scales. The transfer direction of triads of Class-II
depends on the geometry of the three interacting modes while Class III and IV
always transfer energy forward.  In Refs. \cite{Biferale2012prl,Biferale2013jfm} it
was shown that if the NSE is restricted to all homochiral interactions (Class I) it displays a fully isotropic 3D inverse energy cascade.  In
Ref. \cite{sahoo2015role}, a system that transitioned from
the NSE to that of homochiral triads\cite{Biferale2012prl,Biferale2013jfm} was investigated by
introducing a random decimation of modes with negative helicity with a varying
probability, $0 \le \alpha \le 1$ ($\alpha=0$ being the original NSE and $\alpha=1$ being the
system of homochiral triads).  In that study, the
transition from forward to inverse energy cascade happens in a quasi-singular
way such that the inverse cascade exists only at $\alpha \sim 1$ demonstrating
that even if only a small set of interactions among helical waves of both sign
are present (Class II, III and IV), the energy transfer is always forward.
Similar conclusions were reached in Ref. \cite{Kessar2015} where the amplitude of
the negative helical modes was controlled by a dynamical forcing. 

In this work we investigate a variant of the original NSE obtained by
introducing different  weighting of the 4 helical-Fourier classes, such as to
smoothly interpolate from the full NSE to the reduced version \cite{Biferale2012prl,Biferale2013jfm} where
interactions among the  ${\bf u}^+$ and the ${\bf u}^-$ are forbidden, but without removing any modes. In
particular, we evolve the following system:
\begin{equation}                                             
 \partial_t {\bf u} =  
\mathbb{P}\left [\mathcal{N} \right]
-  \nu \Delta^{ 4}   {\bf u}
-  \mu \Delta^{-2}   {\bf u}
+ {\bf F} 
\label{RNS}
\end{equation}
where $\nu$ is the coefficient of the hyperviscosity term and $\mu$ is
the coefficient of the energy sink at large scale needed to arrest the inverse cascade of energy (if
any).  $\mathbb{P}$ is a projection operator to incompressible fields.  The
nonlinearity  $\mathcal{N}$ is defined as
$
 \mathcal{N}=
 \lambda  ({\bf u}   \times {\bf w}  )
 + (1-\lambda) [ \mathbb{P}^+ ({\bf u}^+ \times {\bf w}^+)  
 + \mathbb{P}^- ({\bf u}^- \times {\bf w}^-)],
$ 
where ${\bf w}=\nabla\times {\bf u}$ is the vorticity, and $\mathbb{P}^\pm$ stands
for the projection operator to the incompressible  helical base with ${\bf
u}^\pm=\mathbb{P}^\pm[{\bf u}]$ and $\mathbb{P}=\mathbb{P}^++\mathbb{P}^-$. 
This model, proposed in Ref. \cite{alexakis2016helically}, is graphically summarized
in Fig.~\ref{fig:triad}.  
For any value of $\lambda $ the inviscid system conserves the same quantities
as the 3D NSE,  namely, the energy $E=\frac{1}{2}\langle {\bf u }^2\rangle$ and
 the helicity $H=\frac{1}{2}\langle {\bf u\cdot w} \rangle$ (where 
the angle brackets stand for spatial average), and has the
same rotation, reflection, and dilatation symmetries.
For $\lambda=1$,  $\mathcal{N}$ reduces to the nonlinearity of the NSE and
energy cascades forward.  For $\lambda=0$   the two fields ${\bf u^\pm}$
decouple and Eq.(\ref{RNS}) becomes the equation examined in
Refs. \cite{Biferale2012prl,Biferale2013jfm}.  It conserves two energies
$E^\pm=\frac{1}{2}\langle ({\bf u}^\pm)^2\rangle$ and two sign definite
helicities $H^\pm= \frac{1}{2}\langle {\bf u^\pm \cdot w^\pm } \rangle$
independently and cascades energy inversely. 
We thus expect that as $\lambda$ is varied continuously from $\lambda=1$ to
$\lambda=0$ there will be a change in the direction of energy cascade from
forward to inverse.  The purpose of this work is to investigate how this
transition takes place as the parameter $\lambda$  is varied.    
We perform a systematic series of  high resolution numerical simulations of Eq.
(\ref{RNS}) in a box of size $L=2\pi$.  Energy is
injected  at intermediate wavenumbers $k_f$ by a Gaussian white-in-time forcing 
with a fixed injection rate $\varepsilon_{inj}$.  We use a
pseudo-spectral code, fully dealiased and with second order Adams-Bashforth time
advancing scheme with exact integration of the viscous term.   Table
\ref{param} lists the parameters for all simulations. 
\begin{table}
  \begin{center}
	  \begin{tabular}{c  r  r  c  c  l}
\hline
\hline
  Run  & $\quad N \quad$& $\quad k_f \quad$ &  $\quad \nu\quad $  & $\quad \mu \quad$ & $Re$ \\
\hline
 N1K1  &    $256$         &         $[10,12]$        &   $10^{-14}$      &   $0.5$               &   $  6\times 10^{6}   $ \\
 N2K1  &    $512$         &         $[10,12]$        &   $10^{-16}$      &   $0.5$               &    $ 6\times 10^{8}  $  \\
 N3K1  &   $1024$         &         $[10,12]$        &   $10^{-18}$      &   $0.5$               &     $6\times 10^{10}$   \\
 N2K2  &    $512$         &         $[20,22]$        &   $10^{-16}$      &   $0.5$               &     $5\times 10^{6}$    \\
\hline
\hline
  \end{tabular}
  \caption{Values of the parameters used in the DNS. $N$, spatial resolution; 
  $k_f$, forcing range; $\nu$, viscosity;  $Re=\varepsilon_{inj}^{1/3}/(\nu k_f^{22/3}) $ 
  is the Reynolds number. The large-scale friction $\mu$ is applied for only $ k <k_\mu=2$.}
  \label{param}
  \end{center}
\end{table}
\begin{figure}[!ht]                                                       
  \includegraphics[width=0.49\textwidth]{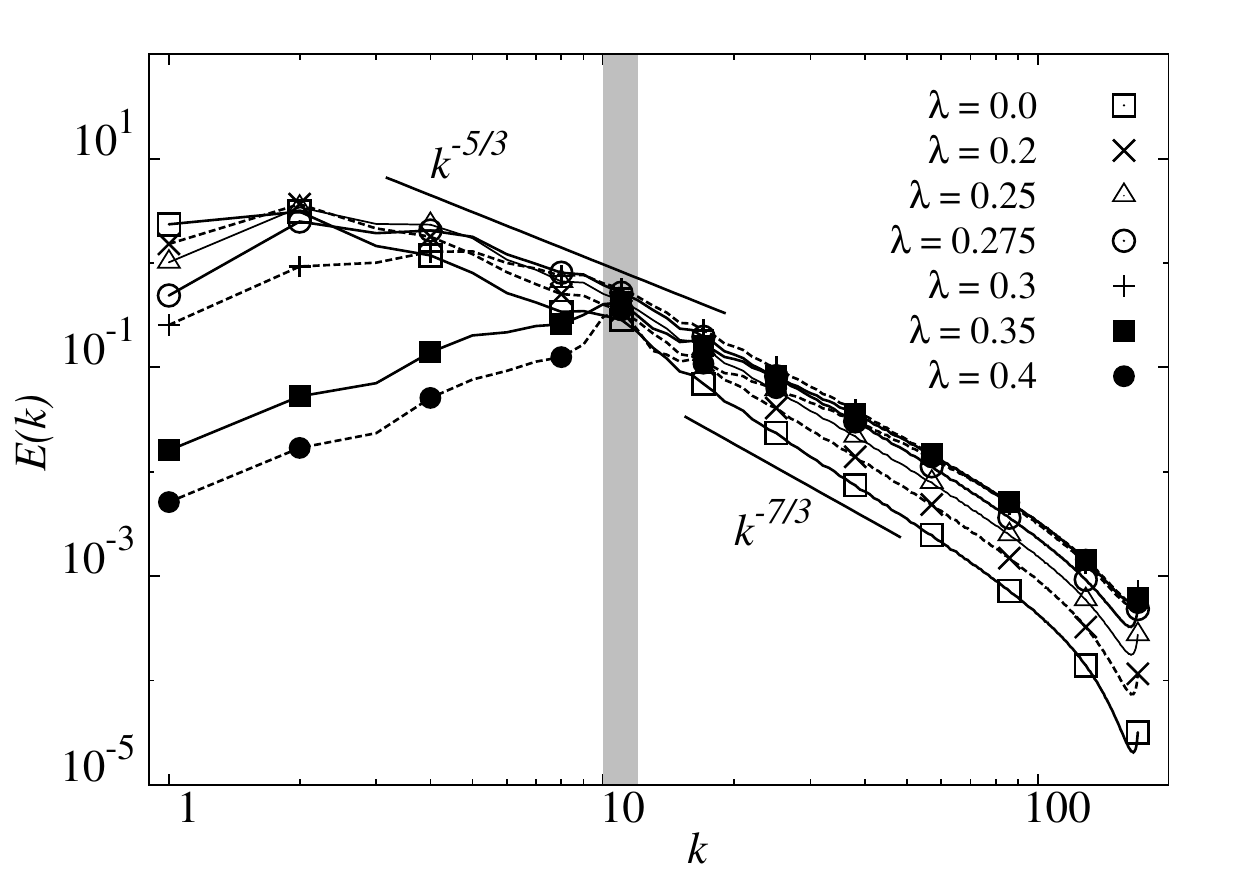}                
   \caption{Log-log plot of energy spectra at changing $\lambda$  
and for fixed forcing range (runs N2K1 in Table I). The gray area denotes  
the forcing window. The two straight lines correspond to the scaling     
predicted in the presence of an energy cascade and for the helicity cascade.     
For $\lambda >\lambda_c \sim 0.3$, there is no inverse energy  
cascade and  $E(k) \propto k^{-5/3}$. For $\lambda < \lambda_c$, we have an  inverse energy transfer 
and a forward  helicity  
transfer, $E(k) \propto k^{-7/3}$. }                                                  
   \label{spec}                                                           
 \end{figure}                                                             
 
 \begin{figure}[!ht]                                               
  \includegraphics[width=0.49\textwidth]{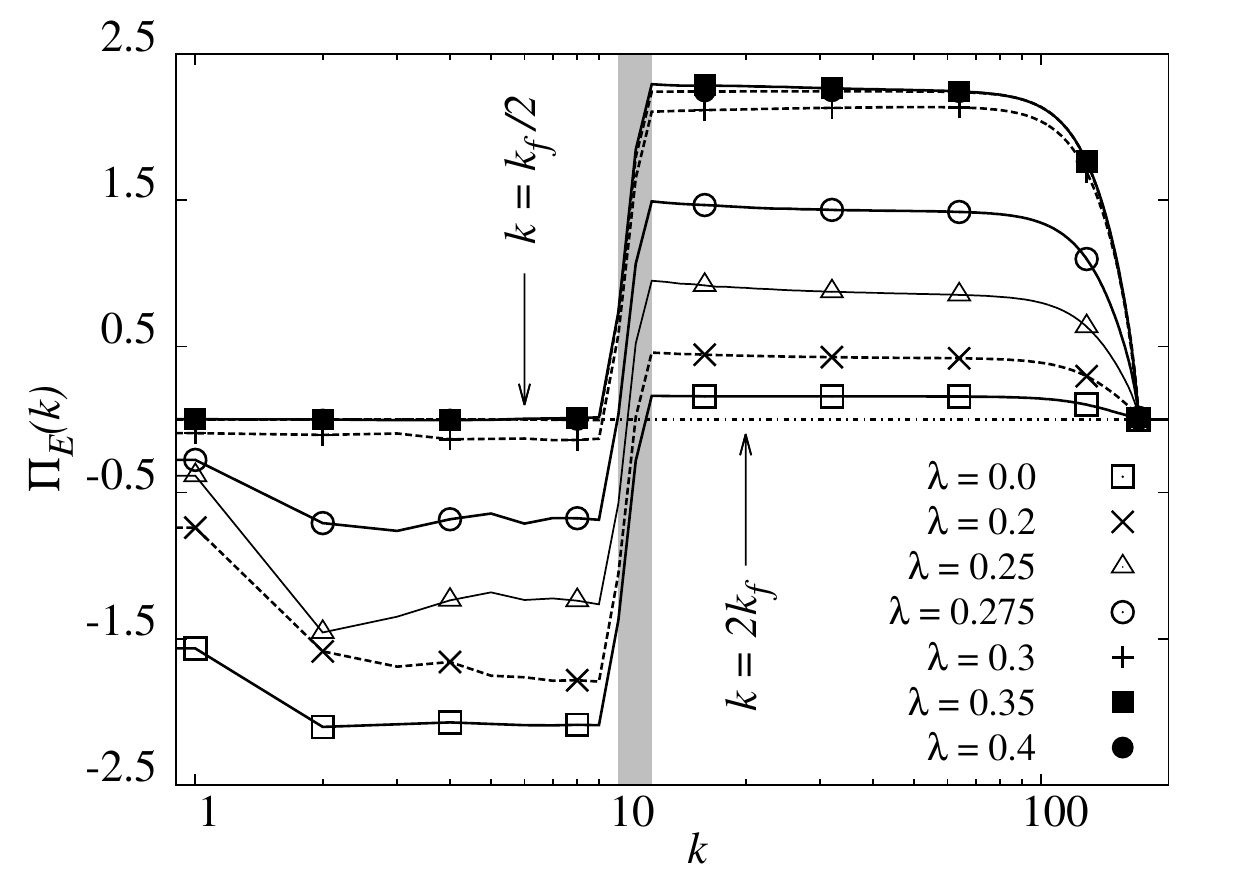}          
   \caption{Energy flux for different values of $\lambda$.
    The gray band shows the forced range of wavenumbers.           
The arrows mark the wavenumbers at which we measure the            
fluctuations in the flux (see insets of Fig.  \ref{inverse}).      
}                                                                  
   \label{flx}                                                     
 \end{figure}                                                      

Figure \ref{spec} shows the energy spectra measured at the steady state for
different values of the parameter $\lambda$ obtained from simulations N2K1.
Clearly, for large values of $\lambda$ there is no significant energy in the
large scales, while small scales display a spectrum compatible with $k^{-5/3}$.
For small values of $\lambda$,  the energy is peaked at large scales forming a
spectrum close to $k^{-5/3}$, while a steeper spectrum closer to $k^{-7/3}$ is
observed in the small scales.  The two behaviors suggest a change from a
forward to an inverse cascade, which is best demonstrated by looking at the
energy fluxes depicted in Fig.~\ref{flx}.  The energy flux is defined as
$                                                                              
\Pi(k) = - \langle {\bf u}^{<}_k \cdot \mathcal{N} \rangle,                       
$                                                                              
where ${\bf u}^{<}_k$ expresses the velocity field ${\bf u}$ filtered so that
its Fourier transform contains only wavenumbers ${\bf k}$ satisfying ${| \bf
k|}\le k$, and expresses that the rate energy is transferred out of the set of
wavenumbers  ${| \bf k|}\le k$ to larger ${\bf k}$.  $\Pi(k)$ is constant in
the inertial ranges $k_\mu\ll k \ll k_f$ and $k_f \ll k \ll k_\nu$ (where
$k_\mu\sim1$ is the hypoviscous wavenumber and $k_\nu\sim N/3$ the
viscous-wavenumber).  It is positive if the cascade is direct  and negative if
the cascade is inverse.

As $\lambda$ is varied the direction of cascade is changing.  For $\lambda \ge
0.3$ the flux is almost zero for $k<k_f$, while it is positive and constant for
$k_f<k<k_\nu$.  For $\lambda \le 0.2$ the opposite picture holds. For $k < k_f$
the flux is negative and constant, while for $k > k_f$ the flux is positive but
weak. For values of $\lambda$ in the range $0.2 < \lambda < 0.3$, we observe a
bidirectional cascade: the coexistence of a forward  and inverse transfer.  Let
us notice that the transition happens  close to  $\lambda=1/3$ that corresponds
to the case where the weight of homochiral triads equals the cumulative weight
of all heterochiral ones.

\begin{figure}[!ht]                                                        
  \includegraphics[width=0.49\textwidth]{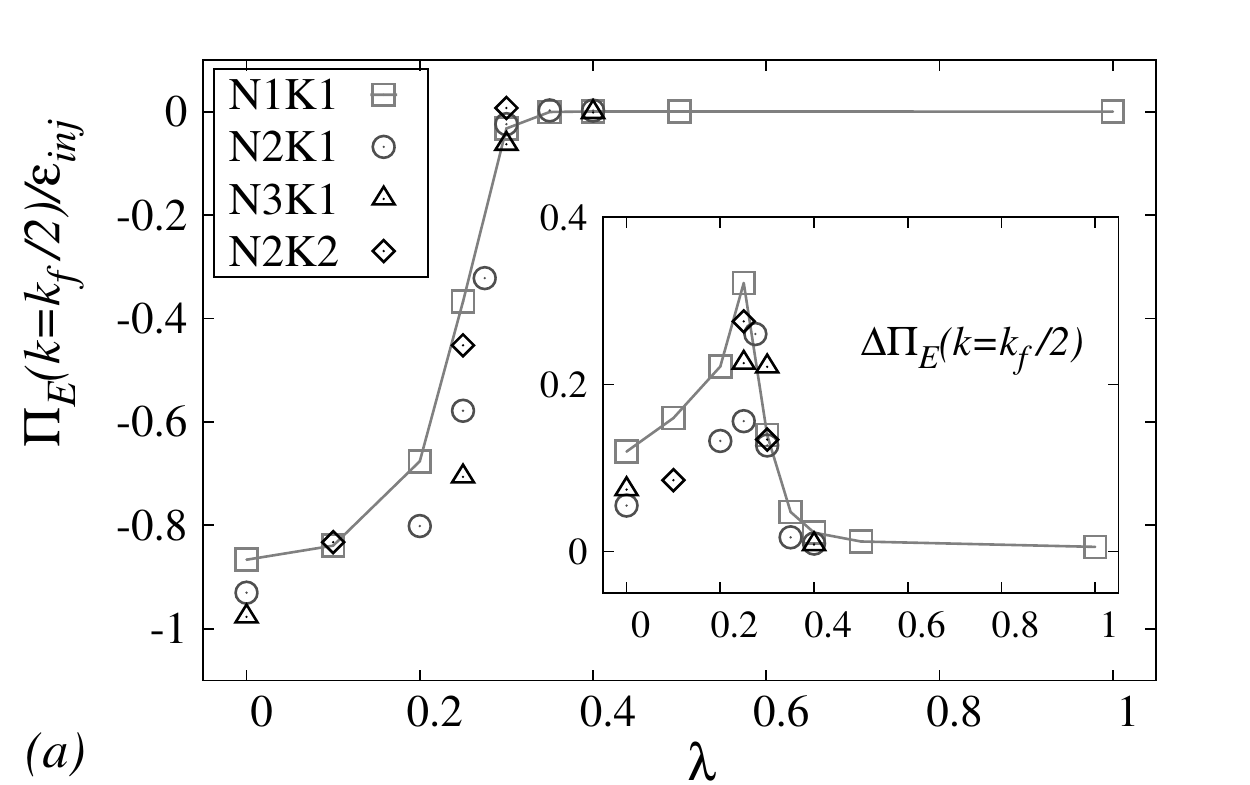}          
 \includegraphics[width=0.49\textwidth]{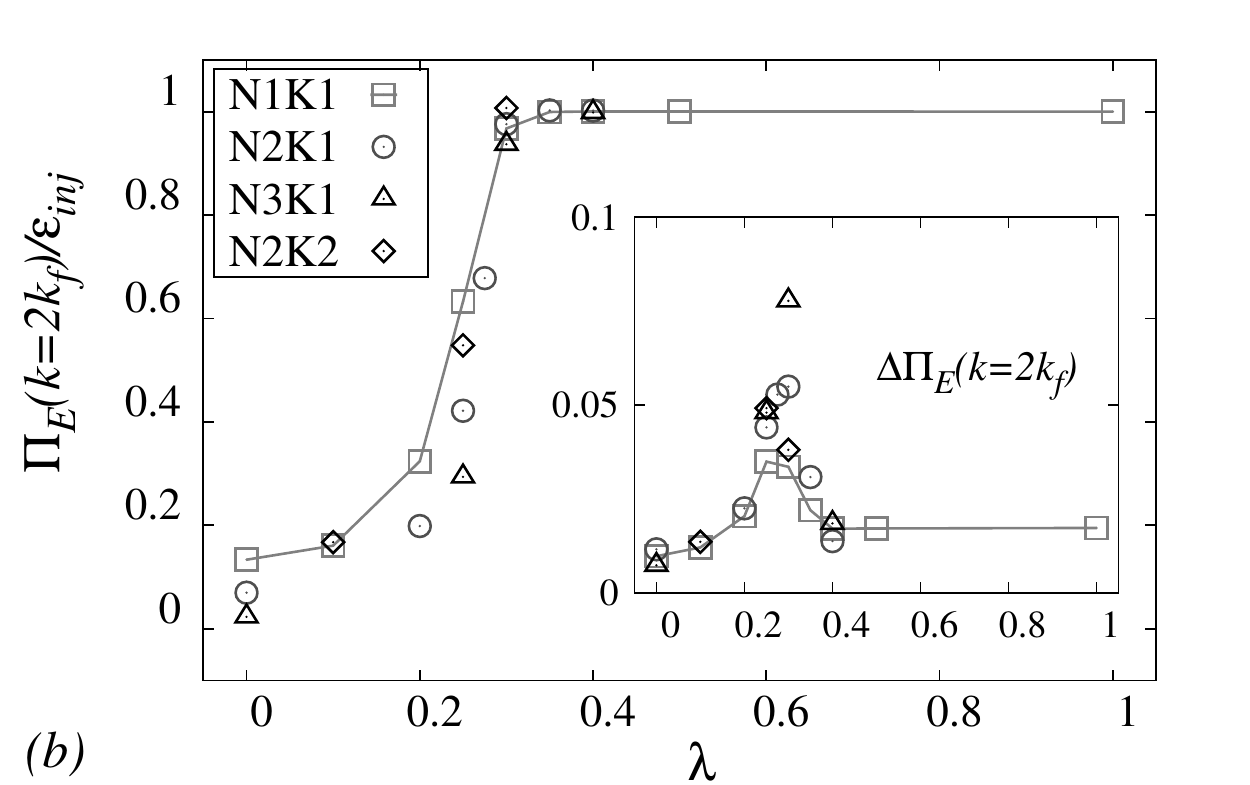}          
   \caption{Normalized energy flux at a scale larger (top) and smaller    
   (bottom) than the forcing range versus $\lambda$.                     
   Insets show the fluctuations around the mean values.                    
  A guiding curve through data points is shown for  N1K1.}                                                                       
   \label{inverse}                                                         
 \end{figure}

The bidirectional cascade is, however, a finite size effect and this behavior
does not survive the large Reynolds number and the large box-size limits, as shown  in Fig. \ref{inverse}.
The inverse flux (measured at the wavenumber $k=k_f/2$) as a function of $\lambda$ 
for different values of the Reynolds numbers (grid sizes) and different box sizes
is shown in Fig.~\ref{inverse}(a) 
while the forward energy flux (measured at the wavenumber $k=2k_f$) is shown in Fig.~\ref{inverse}(b). 
Both fluxes are normalized by the total injection rate $\varepsilon_{inj}$.
The  different symbols correspond to an increase of $Re$   keeping $k_f$ fixed 
(runs N1K1$\to$N2K1$\to$N3K1) or to an increase of $k_f$ keeping $Re$ approximately 
fixed (runs N1K1$\to$N2K2).  For run N1K1 the transition from forward to inverse
cascade is smooth, displaying a bidirectional cascade for values of $\lambda$ in the range  
$0 < \lambda < \lambda_c \simeq 0.3$, while a pure forward cascade 
($|\Pi(k_f/2)|/\varepsilon_{inj} =0$ and $\Pi(2k_f)/\varepsilon_{inj} =1$) is observed for values $\lambda > \lambda_c $.
When $Re$ and $k_f$ are increased, the amplitude of the inverse  cascade 
for the points in the range $0<\lambda < \lambda_c $ is increasing
approaching the value $|\Pi(k_f/2)|/\varepsilon_{inj} =1$, while the forward cascade is decreasing
approaching the value $\Pi(2 k_f)/\varepsilon_{inj} =0$. 
The latter finding suggests that  at infinite $Re$ 
and $k_f$ the cascade is unidirectional and inverse for $\lambda < \lambda_c$, 
while it is unidirectional and forward for $\lambda > \lambda_c$. 
The transition is thus discontinuous. 
This is at difference with  what  observed in quasi-2D systems 
where  the transition occurs in a continuous manner (by a bidirectional cascade) 
similar to a second order phase transition,
and at difference with  what  was observed in Ref. \cite{sahoo2015role} where the transition occurred 
at a singular value of their model parameter, $\alpha \sim 1$.
 
This abrupt transition can be justified by realizing that in a bidirectional
cascade the two inertial ranges ($k_\mu \ll k \ll k_f$ and $k_f   \ll k \ll
k_\nu$) must have different physical properties to sustain different directions
of cascade.  This is possible when, a new dimensional length scale $\ell_*$ is
introduced (e.g., $\ell_*$ is the layer thickness in thin layer turbulence, or
the Zeeman scale in rotating flows) that determines the properties of the flow
due to the external mechanism.  The amplitude of the inverse or forward cascade
then depends on the `distance' of the forcing scale $\ell_f$ from the critical
length scale  $\ell_*$.  In our case, no particular scale $\ell_*$ is
introduced by the parameter $\lambda$.  On the contrary, the inertial ranges
are scale invariant for all values of $\lambda$.  Thus, both  ranges, $\ell
>\ell_f$ and $\ell <\ell_f$,  effectively share the same properties and have to
develop either a forward or a backward cascade, because the flow can not
distinguish the large from the small scales.

 \begin{figure}[!ht]                                                         
  \includegraphics[width=0.49\textwidth]{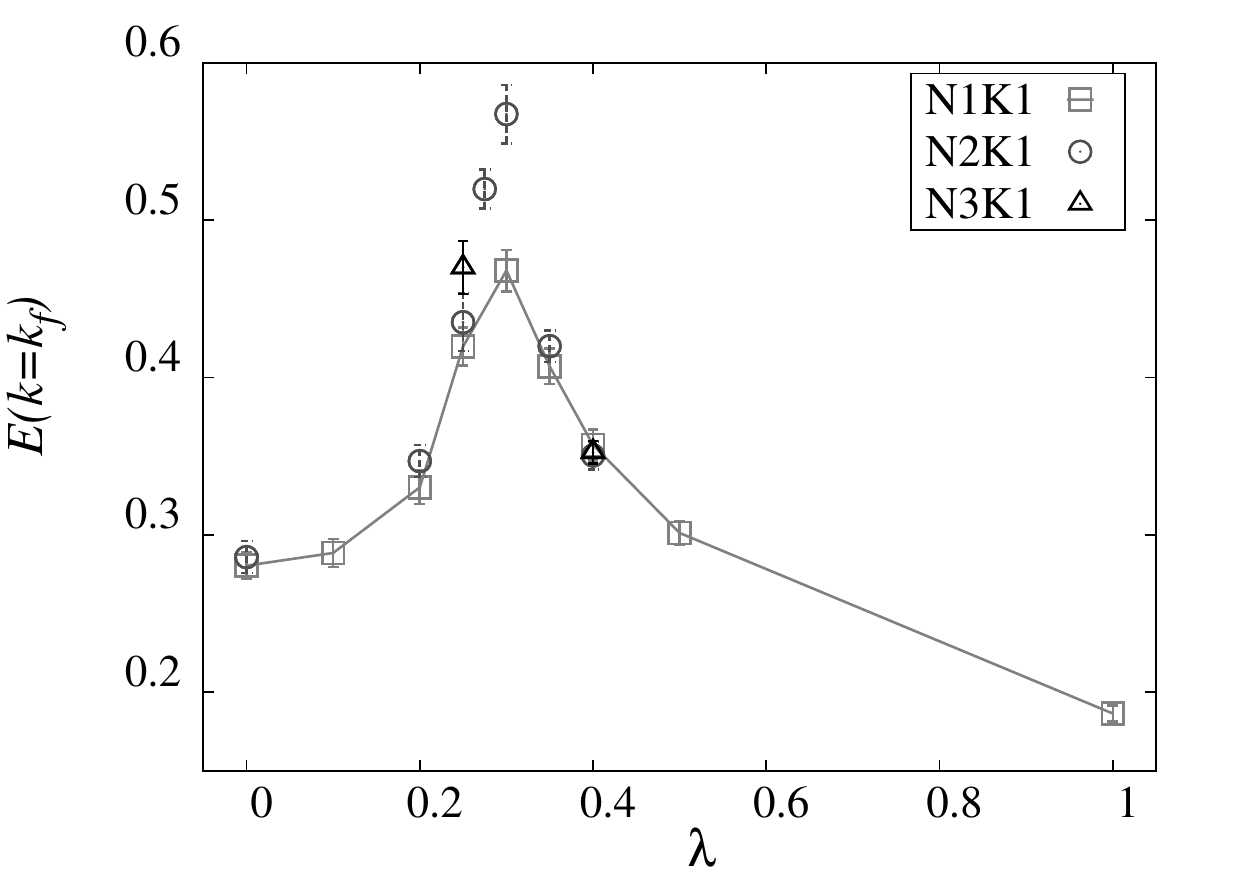}              
   \caption{Energy content at a wavenumber inside the forcing range    
   ($k=11$) versus $\lambda$.
  We plot a guiding curve through data points for  N1K1.}                                                                            
   \label{fig:H/E}                                                           
 \end{figure}                                                                
Not surprisingly, the system displays interesting behavior close to the critical
value $\lambda_c$.  In Fig. ~\ref{fig:H/E} we plot $E(k_f)$, the intensity of
the spectrum at the forcing wavenumbers versus $\lambda$ and for different
Reynolds numbers. The response of the system is critical, showing a tendency
for $E(k_f)$ to diverge as $\lambda\to\lambda_c$.  This divergence is also
reflected in the amplitude of the flux fluctuations $\Delta \Pi$ shown in the
inset of Fig. \ref{inverse} (where $\Delta \Pi$ of run N2K2 is multiplied by
$2^{3/2}$ to account for the $2^{3}$ more interactions involved).  The
existence of multiple phases for the physics of the energy containing eddies is
an important remark that finds support also in recent experimental
empirical findings where turbulent realizations with multiple states have been
observed in swirling and in Taylor-Couette flows \cite{huisman2014multiple,
cortet2010experimental}.

\begin{figure}[!ht]
    \centering
        \includegraphics[width=\columnwidth]{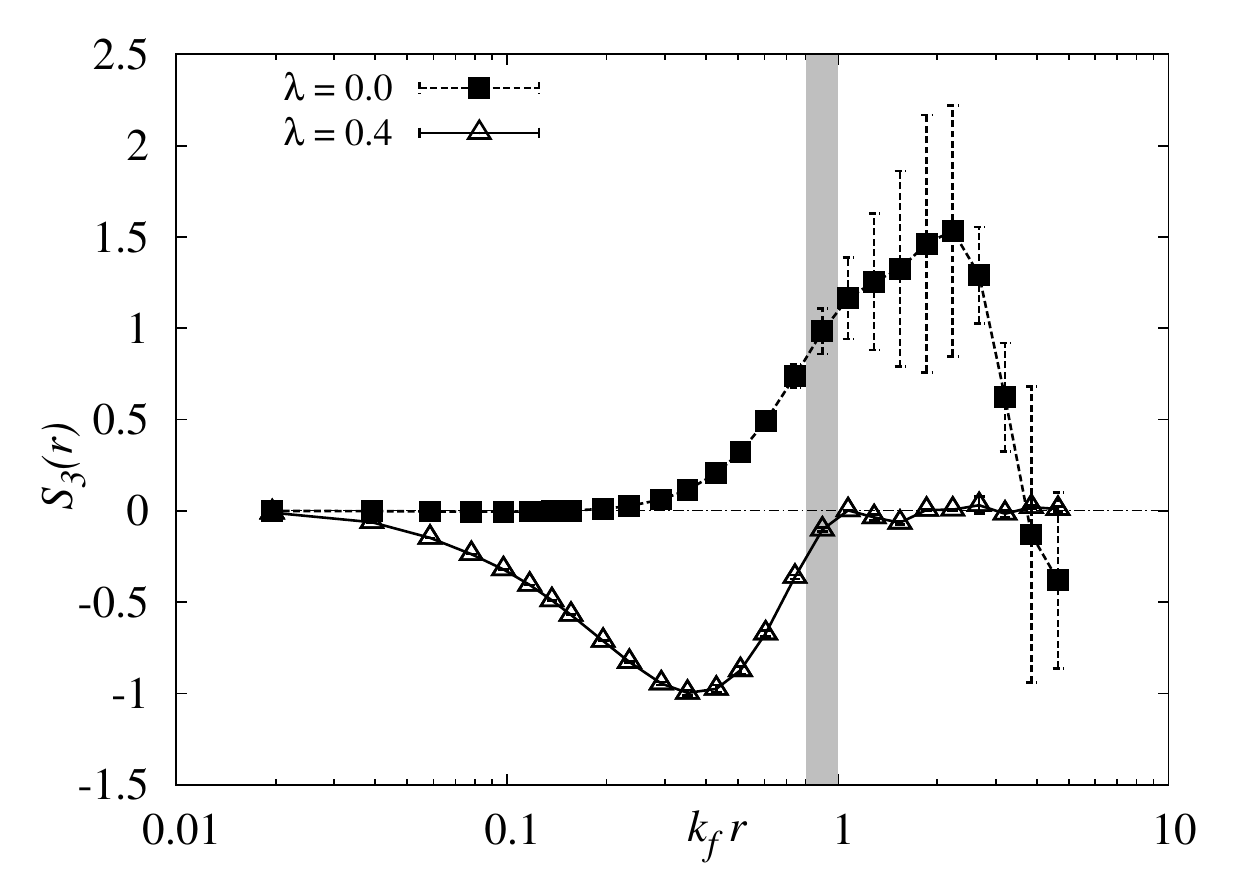}
  \caption{Third order structure functions for the two  cases with only direct or inverse energy cascade $\lambda=0.4$ and $\lambda=0$.}
  \label{fig:sf3}
\end{figure}

The direction of the energy transfer can also be studied by looking at the
behavior of the structure functions $S_n(r) = \langle (\delta {\bf u}_\|(r) )^n
\rangle $, where $\delta {\bf u}_\| (r)= ({\bf u(x+r)-u(x)})\cdot {\bf r}/r $,
that have the advantage of being easily measured in
experiments. In particular, for the original NSE,  the von K\'arm\'an-Howarth
equation states that the third order  structure function is related to the
direction of the cascade and it is negative for a forward transfer and positive
for a backward transfer. In the form of the NSE
investigated here [Eq. (\ref{RNS})], the von K\'arm\'an-Howarth equation is more
complicated (see, e.g., Appendix A.1 of Ref. \cite{Biferale2013jfm} for the case with
$\lambda=0$). Nevertheless, we show in Fig.~\ref{fig:sf3} that even a simple
measurement based on $S_3(r)$ is in good agreement with the indication that for
$r > r_f=2\pi/k_f$ the sign does change by crossing $\lambda_c$. 

In this work we have  demonstrated that by controlling the amplitude of the
interactions in the NSE the energy cascade can change
direction from forward to inverse and \emph{vice versa}. 
In the model used here, this change of direction is not due to previously known
mechanisms, e.g.,  a change in the dimensionality, a change in the ideal
invariants, or the breaking of any symmetry of the original equations caused by
the introduction of external forcing as in the presence of rotation or of a
magnetic field,
revealing that the fully nonlinear dynamics of the 3D NSE
is more complex than what was told by the accepted phenomenology.
In particular, we showed that the energy cascade is strongly sensitive to
the relative dynamical weight of homochiral to heterochiral helical Fourier
interactions, suggesting the search for similar footprints of inverse energy
transfer also in other empirical turbulent realization.

The mechanisms revealed here could be relevant to physical systems.
In the case of rotation, the eigenmodes of the linear operator
are in fact the helical modes used here, with the different sign helical modes
having opposite direction of propagation. It is thus possible (although it still
needs to be shown) that opposite helicity modes decorrelate faster and the
relevant nonlinearities quench faster than same helicity modes. Similar
properties might be at play in magnetohydrodynamics and in active fluids \cite{linkmann,dunkel}.

Our results indicate that the transition becomes discontinuous in the large
$Re$ limit.  This is the first time that such a discontinuous transition has been
reported for the cascade direction.
We have linked this discontinuity of the transition with the
preservation of scale similarity in our model; thus, the same arguments can
also be applied to other out-of-equilibrium systems with scale similarity that do not
originate from the NSE. 

The presence of a control parameter in the
turbulence model is key also to validate or benchmark the analytical theory of
turbulence, e.g., renormalization group approaches or closures
\cite{canet,giles,lvov,antonov}.
Our work thus points to a new direction in which the NSE (for $\lambda=1$)
can be viewed as a system `close' to criticality (for which
$\lambda=\lambda_c$)  that can lead to new theoretical investigations
in strongly out-of-equilibrium statistical mechanics.

\begin{acknowledgements} 
The research leading to these results has received
funding from the European Union's Seventh Framework Programme (FP7/2007-2013)
under grant agreement No.  339032.  
\end{acknowledgements}

\bibliography{references}

\begin{thebibliography}{34}
\expandafter\ifx\csname natexlab\endcsname\relax\def\natexlab#1{#1}\fi
\expandafter\ifx\csname bibnamefont\endcsname\relax
  \def\bibnamefont#1{#1}\fi
\expandafter\ifx\csname bibfnamefont\endcsname\relax
  \def\bibfnamefont#1{#1}\fi
\expandafter\ifx\csname citenamefont\endcsname\relax
  \def\citenamefont#1{#1}\fi
\expandafter\ifx\csname url\endcsname\relax
  \def\url#1{\texttt{#1}}\fi
\expandafter\ifx\csname urlprefix\endcsname\relax\def\urlprefix{URL }\fi
\providecommand{\bibinfo}[2]{#2}
\providecommand{\eprint}[2][]{\url{#2}}

\bibitem[{\citenamefont{Frisch}(1995)}]{Frisch}
\bibinfo{author}{\bibfnamefont{U.}~\bibnamefont{Frisch}},
  \emph{\bibinfo{title}{Turbulence: The Legacy of A. N. Kolmogorov}}
  (\bibinfo{publisher}{Cambridge University Press, Cambridge, England},
  \bibinfo{year}{1995}).

\bibitem[{\citenamefont{Brissaud et~al.}(1973)\citenamefont{Brissaud, Frisch,
  Leorat, Lesieur, and Mazure}}]{brissaud1973helicity}
\bibinfo{author}{\bibfnamefont{A.}~\bibnamefont{Brissaud}},
  \bibinfo{author}{\bibfnamefont{U.}~\bibnamefont{Frisch}},
  \bibinfo{author}{\bibfnamefont{J.}~\bibnamefont{Leorat}},
  \bibinfo{author}{\bibfnamefont{M.}~\bibnamefont{Lesieur}}, \bibnamefont{and}
  \bibinfo{author}{\bibfnamefont{A.}~\bibnamefont{Mazure}},
  \bibinfo{journal}{Phys. Fluids} \textbf{\bibinfo{volume}{16}},
  \bibinfo{pages}{1366} (\bibinfo{year}{1973}).

\bibitem[{\citenamefont{Chen et~al.}(2003)\citenamefont{Chen, Chen, and
  Eyink}}]{chen2003joint}
\bibinfo{author}{\bibfnamefont{Q.}~\bibnamefont{Chen}},
  \bibinfo{author}{\bibfnamefont{S.}~\bibnamefont{Chen}}, \bibnamefont{and}
  \bibinfo{author}{\bibfnamefont{G.~L.} \bibnamefont{Eyink}},
  \bibinfo{journal}{Phys. Fluids} \textbf{\bibinfo{volume}{15}},
  \bibinfo{pages}{361} (\bibinfo{year}{2003}).

\bibitem[{\citenamefont{{Smith} and {Waleffe}}(1999)}]{Smith1999transfer}
\bibinfo{author}{\bibfnamefont{L.~M.} \bibnamefont{{Smith}}} \bibnamefont{and}
  \bibinfo{author}{\bibfnamefont{F.}~\bibnamefont{{Waleffe}}},
  \bibinfo{journal}{Phys. Fluids} \textbf{\bibinfo{volume}{11}},
  \bibinfo{pages}{1608} (\bibinfo{year}{1999}).

\bibitem[{\citenamefont{{Celani} et~al.}(2010)\citenamefont{{Celani},
  {Musacchio}, and {Vincenzi}}}]{celani2010turbulence}
\bibinfo{author}{\bibfnamefont{A.}~\bibnamefont{{Celani}}},
  \bibinfo{author}{\bibfnamefont{S.}~\bibnamefont{{Musacchio}}},
  \bibnamefont{and}
  \bibinfo{author}{\bibfnamefont{D.}~\bibnamefont{{Vincenzi}}},
  \bibinfo{journal}{Phys. Rev. Lett.} \textbf{\bibinfo{volume}{104}},
  \bibinfo{eid}{184506} (\bibinfo{year}{2010}).

\bibitem[{\citenamefont{{Alexakis}}(2011)}]{alexakis2011two}
\bibinfo{author}{\bibfnamefont{A.}~\bibnamefont{{Alexakis}}},
  \bibinfo{journal}{Phys. Rev. E} \textbf{\bibinfo{volume}{84}},
  \bibinfo{eid}{056330} (\bibinfo{year}{2011}).

\bibitem[{\citenamefont{{Pouquet} and {Marino}}(2013)}]{pouquet2013geophysical}
\bibinfo{author}{\bibfnamefont{A.}~\bibnamefont{{Pouquet}}} \bibnamefont{and}
  \bibinfo{author}{\bibfnamefont{R.}~\bibnamefont{{Marino}}},
  \bibinfo{journal}{Phys. Rev. Lett.} \textbf{\bibinfo{volume}{111}},
  \bibinfo{pages}{234501} (\bibinfo{year}{2013}).

\bibitem[{\citenamefont{{Marino} et~al.}(2013)\citenamefont{{Marino},
  {Mininni}, {Rosenberg}, and {Pouquet}}}]{marino2013inverse}
\bibinfo{author}{\bibfnamefont{R.}~\bibnamefont{{Marino}}},
  \bibinfo{author}{\bibfnamefont{P.~D.} \bibnamefont{{Mininni}}},
  \bibinfo{author}{\bibfnamefont{D.}~\bibnamefont{{Rosenberg}}},
  \bibnamefont{and}
  \bibinfo{author}{\bibfnamefont{A.}~\bibnamefont{{Pouquet}}},
  \bibinfo{journal}{Europhys. Lett.} \textbf{\bibinfo{volume}{102}},
  \bibinfo{eid}{44006} (\bibinfo{year}{2013}).

\bibitem[{\citenamefont{Deusebio et~al.}(2014)\citenamefont{Deusebio, Boffetta,
  Lindborg, and Musacchio}}]{deusebio2014dimensional}
\bibinfo{author}{\bibfnamefont{E.}~\bibnamefont{Deusebio}},
  \bibinfo{author}{\bibfnamefont{G.}~\bibnamefont{Boffetta}},
  \bibinfo{author}{\bibfnamefont{E.}~\bibnamefont{Lindborg}}, \bibnamefont{and}
  \bibinfo{author}{\bibfnamefont{S.}~\bibnamefont{Musacchio}},
  \bibinfo{journal}{Phys. Rev. E} \textbf{\bibinfo{volume}{90}},
  \bibinfo{pages}{023005} (\bibinfo{year}{2014}).

\bibitem[{\citenamefont{Seshasayanan et~al.}(2014)\citenamefont{Seshasayanan,
  Benavides, and Alexakis}}]{seshasayanan2014on}
\bibinfo{author}{\bibfnamefont{K.}~\bibnamefont{Seshasayanan}},
  \bibinfo{author}{\bibfnamefont{S.~J.} \bibnamefont{Benavides}},
  \bibnamefont{and} \bibinfo{author}{\bibfnamefont{A.}~\bibnamefont{Alexakis}},
  \bibinfo{journal}{Phys. Rev. E} \textbf{\bibinfo{volume}{90}},
  \bibinfo{pages}{051003} (\bibinfo{year}{2014}).

\bibitem[{\citenamefont{Seshasayanan and
  Alexakis}(2016)}]{seshasayanan2016critical}
\bibinfo{author}{\bibfnamefont{K.}~\bibnamefont{Seshasayanan}}
  \bibnamefont{and} \bibinfo{author}{\bibfnamefont{A.}~\bibnamefont{Alexakis}},
  \bibinfo{journal}{Phys. Rev. E} \textbf{\bibinfo{volume}{93}},
  \bibinfo{pages}{013104} (\bibinfo{year}{2016}).

\bibitem[{\citenamefont{{Benavides} and {Alexakis}}(2017)}]{Benavides2017}
\bibinfo{author}{\bibfnamefont{S.~J.} \bibnamefont{{Benavides}}}
  \bibnamefont{and}
  \bibinfo{author}{\bibfnamefont{A.}~\bibnamefont{{Alexakis}}},
  \bibinfo{journal}{ArXiv e-prints}  (\bibinfo{year}{2017}),
  \eprint{arXiv:1701.05162}.

\bibitem[{\citenamefont{Biferale et~al.}(2016)\citenamefont{Biferale,
  Bonaccorso, Mazzitelli, van Hinsberg, Lanotte, Musacchio, Perlekar, and
  Toschi}}]{Biferale2016}
\bibinfo{author}{\bibfnamefont{L.}~\bibnamefont{Biferale}},
  \bibinfo{author}{\bibfnamefont{F.}~\bibnamefont{Bonaccorso}},
  \bibinfo{author}{\bibfnamefont{I.~M.} \bibnamefont{Mazzitelli}},
  \bibinfo{author}{\bibfnamefont{M.~A.~T.} \bibnamefont{van Hinsberg}},
  \bibinfo{author}{\bibfnamefont{A.~S.} \bibnamefont{Lanotte}},
  \bibinfo{author}{\bibfnamefont{S.}~\bibnamefont{Musacchio}},
  \bibinfo{author}{\bibfnamefont{P.}~\bibnamefont{Perlekar}}, \bibnamefont{and}
  \bibinfo{author}{\bibfnamefont{F.}~\bibnamefont{Toschi}},
  \bibinfo{journal}{Phys. Rev. X} \textbf{\bibinfo{volume}{6}},
  \bibinfo{pages}{041036} (\bibinfo{year}{2016}).

\bibitem[{\citenamefont{{Xia} et~al.}(2011)\citenamefont{{Xia}, {Byrne},
  {Falkovich}, and {Shats}}}]{xia2011upscaling}
\bibinfo{author}{\bibfnamefont{H.}~\bibnamefont{{Xia}}},
  \bibinfo{author}{\bibfnamefont{D.}~\bibnamefont{{Byrne}}},
  \bibinfo{author}{\bibfnamefont{G.}~\bibnamefont{{Falkovich}}},
  \bibnamefont{and} \bibinfo{author}{\bibfnamefont{M.}~\bibnamefont{{Shats}}},
  \bibinfo{journal}{Nature Physics} \textbf{\bibinfo{volume}{7}},
  \bibinfo{pages}{321} (\bibinfo{year}{2011}).

\bibitem[{\citenamefont{{Yarom} et~al.}(2013)\citenamefont{{Yarom}, {Vardi},
  and {Sharon}}}]{Yarom2013}
\bibinfo{author}{\bibfnamefont{E.}~\bibnamefont{{Yarom}}},
  \bibinfo{author}{\bibfnamefont{Y.}~\bibnamefont{{Vardi}}}, \bibnamefont{and}
  \bibinfo{author}{\bibfnamefont{E.}~\bibnamefont{{Sharon}}},
  \bibinfo{journal}{Phys. Fluids} \textbf{\bibinfo{volume}{25}},
  \bibinfo{pages}{085105} (\bibinfo{year}{2013}).

\bibitem[{\citenamefont{{Campagne} et~al.}(2014)\citenamefont{{Campagne},
  {Gallet}, {Moisy}, and {Cortet}}}]{campagne2014direct}
\bibinfo{author}{\bibfnamefont{A.}~\bibnamefont{{Campagne}}},
  \bibinfo{author}{\bibfnamefont{B.}~\bibnamefont{{Gallet}}},
  \bibinfo{author}{\bibfnamefont{F.}~\bibnamefont{{Moisy}}}, \bibnamefont{and}
  \bibinfo{author}{\bibfnamefont{P.-P.} \bibnamefont{{Cortet}}},
  \bibinfo{journal}{Phys. Fluids} \textbf{\bibinfo{volume}{26}},
  \bibinfo{eid}{125112} (\bibinfo{year}{2014}).

\bibitem[{\citenamefont{{Byrne} and {Zhang}}(2013)}]{huricanes}
\bibinfo{author}{\bibfnamefont{D.}~\bibnamefont{{Byrne}}} \bibnamefont{and}
  \bibinfo{author}{\bibfnamefont{J.~A.} \bibnamefont{{Zhang}}},
  \bibinfo{journal}{Geophys. Res. Lett.} \textbf{\bibinfo{volume}{40}},
  \bibinfo{pages}{1439} (\bibinfo{year}{2013}).

\bibitem[{\citenamefont{Craya}(1958)}]{craya1958contributiona}
\bibinfo{author}{\bibfnamefont{A.}~\bibnamefont{Craya}}, \bibinfo{journal}{Sci.
  Tech. du Ministere de l’Air (France)}  (\bibinfo{year}{1958}).

\bibitem[{\citenamefont{Herring et~al.}(1982)\citenamefont{Herring, Schertzer,
  Lesieur, Newman, Chollet, and Larcheveque}}]{herring1982comparative}
\bibinfo{author}{\bibfnamefont{J.}~\bibnamefont{Herring}},
  \bibinfo{author}{\bibfnamefont{D.}~\bibnamefont{Schertzer}},
  \bibinfo{author}{\bibfnamefont{M.}~\bibnamefont{Lesieur}},
  \bibinfo{author}{\bibfnamefont{G.}~\bibnamefont{Newman}},
  \bibinfo{author}{\bibfnamefont{J.}~\bibnamefont{Chollet}}, \bibnamefont{and}
  \bibinfo{author}{\bibfnamefont{M.}~\bibnamefont{Larcheveque}},
  \bibinfo{journal}{J. Fluid Mech.} \textbf{\bibinfo{volume}{124}},
  \bibinfo{pages}{411} (\bibinfo{year}{1982}).

\bibitem[{\citenamefont{Lesieur}(2008)}]{Lesieur72}
\bibinfo{author}{\bibfnamefont{M.}~\bibnamefont{Lesieur}},
  \emph{\bibinfo{title}{Turbulence in Fluids}} (\bibinfo{publisher}{Springer,
  The Netherlands}, \bibinfo{year}{2008}).

\bibitem[{\citenamefont{{Waleffe}}(1992)}]{waleffe1992nature}
\bibinfo{author}{\bibfnamefont{F.}~\bibnamefont{{Waleffe}}},
  \bibinfo{journal}{Phys. Fluids} \textbf{\bibinfo{volume}{4}},
  \bibinfo{pages}{350} (\bibinfo{year}{1992}).

\bibitem[{\citenamefont{{Biferale} et~al.}(2012)\citenamefont{{Biferale},
  {Musacchio}, and {Toschi}}}]{Biferale2012prl}
\bibinfo{author}{\bibfnamefont{L.}~\bibnamefont{{Biferale}}},
  \bibinfo{author}{\bibfnamefont{S.}~\bibnamefont{{Musacchio}}},
  \bibnamefont{and} \bibinfo{author}{\bibfnamefont{F.}~\bibnamefont{{Toschi}}},
  \bibinfo{journal}{Phys. Rev. Lett.} \textbf{\bibinfo{volume}{108}},
  \bibinfo{pages}{164501} (\bibinfo{year}{2012}).

\bibitem[{\citenamefont{{Biferale} et~al.}(2013)\citenamefont{{Biferale},
  {Musacchio}, and {Toschi}}}]{Biferale2013jfm}
\bibinfo{author}{\bibfnamefont{L.}~\bibnamefont{{Biferale}}},
  \bibinfo{author}{\bibfnamefont{S.}~\bibnamefont{{Musacchio}}},
  \bibnamefont{and} \bibinfo{author}{\bibfnamefont{F.}~\bibnamefont{{Toschi}}},
  \bibinfo{journal}{J. Fluid Mech.} \textbf{\bibinfo{volume}{730}},
  \bibinfo{pages}{309} (\bibinfo{year}{2013}).

\bibitem[{\citenamefont{Sahoo et~al.}(2015)\citenamefont{Sahoo, Bonaccorso, and
  Biferale}}]{sahoo2015role}
\bibinfo{author}{\bibfnamefont{G.}~\bibnamefont{Sahoo}},
  \bibinfo{author}{\bibfnamefont{F.}~\bibnamefont{Bonaccorso}},
  \bibnamefont{and} \bibinfo{author}{\bibfnamefont{L.}~\bibnamefont{Biferale}},
  \bibinfo{journal}{Phys. Rev. E} \textbf{\bibinfo{volume}{92}},
  \bibinfo{pages}{051002} (\bibinfo{year}{2015}).

\bibitem[{\citenamefont{Kessar et~al.}(2015)\citenamefont{Kessar, Plunian,
  Stepanov, and Balarac}}]{Kessar2015}
\bibinfo{author}{\bibfnamefont{M.}~\bibnamefont{Kessar}},
  \bibinfo{author}{\bibfnamefont{F.}~\bibnamefont{Plunian}},
  \bibinfo{author}{\bibfnamefont{R.}~\bibnamefont{Stepanov}}, \bibnamefont{and}
  \bibinfo{author}{\bibfnamefont{G.}~\bibnamefont{Balarac}},
  \bibinfo{journal}{Phys. Rev. E} \textbf{\bibinfo{volume}{92}},
  \bibinfo{pages}{031004} (\bibinfo{year}{2015}).

\bibitem[{\citenamefont{Alexakis}(2017)}]{alexakis2016helically}
\bibinfo{author}{\bibfnamefont{A.}~\bibnamefont{Alexakis}},
  \bibinfo{journal}{J. Fluid Mech.} \textbf{\bibinfo{volume}{812}},
  \bibinfo{pages}{752} (\bibinfo{year}{2017}).

\bibitem[{\citenamefont{Huisman et~al.}(2014)\citenamefont{Huisman, Van
  Der~Veen, Sun, and Lohse}}]{huisman2014multiple}
\bibinfo{author}{\bibfnamefont{S.~G.} \bibnamefont{Huisman}},
  \bibinfo{author}{\bibfnamefont{R.~C.} \bibnamefont{Van Der~Veen}},
  \bibinfo{author}{\bibfnamefont{C.}~\bibnamefont{Sun}}, \bibnamefont{and}
  \bibinfo{author}{\bibfnamefont{D.}~\bibnamefont{Lohse}},
  \bibinfo{journal}{Nat. Commun.} \textbf{\bibinfo{volume}{5}}
  (\bibinfo{year}{2014}).

\bibitem[{\citenamefont{Cortet et~al.}(2010)\citenamefont{Cortet, Chiffaudel,
  Daviaud, and Dubrulle}}]{cortet2010experimental}
\bibinfo{author}{\bibfnamefont{P.-P.} \bibnamefont{Cortet}},
  \bibinfo{author}{\bibfnamefont{A.}~\bibnamefont{Chiffaudel}},
  \bibinfo{author}{\bibfnamefont{F.}~\bibnamefont{Daviaud}}, \bibnamefont{and}
  \bibinfo{author}{\bibfnamefont{B.}~\bibnamefont{Dubrulle}},
  \bibinfo{journal}{Phys. Rev. Lett.} \textbf{\bibinfo{volume}{105}},
  \bibinfo{pages}{214501} (\bibinfo{year}{2010}).

\bibitem[{\citenamefont{Linkmann et~al.}(2017)\citenamefont{Linkmann, Sahoo,
  McKay, Berera, and Biferale}}]{linkmann}
\bibinfo{author}{\bibfnamefont{M.}~\bibnamefont{Linkmann}},
  \bibinfo{author}{\bibfnamefont{G.}~\bibnamefont{Sahoo}},
  \bibinfo{author}{\bibfnamefont{M.}~\bibnamefont{McKay}},
  \bibinfo{author}{\bibfnamefont{A.}~\bibnamefont{Berera}}, \bibnamefont{and}
  \bibinfo{author}{\bibfnamefont{L.}~\bibnamefont{Biferale}},
  \bibinfo{journal}{Astrophys. J.} \textbf{\bibinfo{volume}{836}},
  \bibinfo{pages}{26} (\bibinfo{year}{2017}).

\bibitem[{\citenamefont{Stomka and Dunkel}(2017)}]{dunkel}
\bibinfo{author}{\bibfnamefont{J.}~\bibnamefont{Stomka}} \bibnamefont{and}
  \bibinfo{author}{\bibfnamefont{J.}~\bibnamefont{Dunkel}},
  \bibinfo{journal}{Proc. of the Natl. Acad. Sci.}
  \textbf{\bibinfo{volume}{114}}, \bibinfo{pages}{2119} (\bibinfo{year}{2017}).

\bibitem[{\citenamefont{Canet et~al.}(2016)\citenamefont{Canet, Delamotte, and
  Wschebor}}]{canet}
\bibinfo{author}{\bibfnamefont{L.}~\bibnamefont{Canet}},
  \bibinfo{author}{\bibfnamefont{B.}~\bibnamefont{Delamotte}},
  \bibnamefont{and} \bibinfo{author}{\bibfnamefont{N.}~\bibnamefont{Wschebor}},
  \bibinfo{journal}{Phys. Rev. E} \textbf{\bibinfo{volume}{93}},
  \bibinfo{pages}{063101} (\bibinfo{year}{2016}).

\bibitem[{\citenamefont{Giles}(2001)}]{giles}
\bibinfo{author}{\bibfnamefont{M.~J.} \bibnamefont{Giles}},
  \bibinfo{journal}{J. Phys. A: Mathematical and General}
  \textbf{\bibinfo{volume}{34}}, \bibinfo{pages}{4389} (\bibinfo{year}{2001}).

\bibitem[{\citenamefont{L'vov and Procaccia}(2000)}]{lvov}
\bibinfo{author}{\bibfnamefont{V.~S.} \bibnamefont{L'vov}} \bibnamefont{and}
  \bibinfo{author}{\bibfnamefont{I.}~\bibnamefont{Procaccia}},
  \bibinfo{journal}{Phys. Rev. E} \textbf{\bibinfo{volume}{62}},
  \bibinfo{pages}{8037} (\bibinfo{year}{2000}).

\bibitem[{\citenamefont{Adzhemyan et~al.}(1999)\citenamefont{Adzhemyan,
  Antonov, and Vasil'ev}}]{antonov}
\bibinfo{author}{\bibfnamefont{L.~T.} \bibnamefont{Adzhemyan}},
  \bibinfo{author}{\bibfnamefont{N.~V.} \bibnamefont{Antonov}},
  \bibnamefont{and} \bibinfo{author}{\bibfnamefont{A.~N.}
  \bibnamefont{Vasil'ev}}, \emph{\bibinfo{title}{The Field Theoretic
  Renormalization Group in Fully Developed Turbulence}}
  (\bibinfo{publisher}{Gordon \& Breach, London}, \bibinfo{year}{1999}).

\end{thebibliography}
\end{document}